\documentclass[a4paper]{cas-dc}

\usepackage[numbers]{natbib}
\usepackage{verbatim}

\usepackage{amsmath} % For align, split, \underbrace, \text
\usepackage{amssymb} % For \agt
\usepackage{mathtools} % For \agt

\begin{document}
\let\WriteBookmarks\relax
\def\floatpagepagefraction{1}
\def\textpagefraction{.001}
\shorttitle{}
\shortauthors{H.B. Schonfeld et~al.}

\title [mode = title]{Closing the ultrahigh temperature metrology gap: non-contact thermal conductivity ($\mathrm{k}$) and spectral emittance ($\mathrm{\varepsilon_{\lambda}}$) of molybdenum up to 3200 K}                      
%\tnotemark[1,2]

\author[1]{Hunter B. Schonfeld}[type=editor,
                        orcid=0009-0003-1777-6382]
%\cormark[1]
\ead{kxs7pc@virginia.edu}
\cormark[1]

\credit{Writing – original draft, Conceptualization, Data curation, Formal Analysis, Investigation, Methodology, Validation, Visualization}

%\address[1]{, Street 129, 1043 NX Amsterdam, The Netherlands}
\affiliation[1]{organization={Department of Mechanical and
Aerospace Engineering, University of Virginia}, 
                city={Charlottesville},
%               citysep={}, % Uncomment if no comma needed between city and postcode
                postcode={22904}, 
                state={VA},
                country={USA}}

\author[1]{Elizabeth Golightly}

\credit{Investigation}

\author[1]{Milena Milich}

\credit{Conceptualization, Validation}

\author[1]{Scott Bender}

\credit{Formal Analysis}

\author[2]{Konstantinos Boboridis}[type=editor,
                        orcid=0000-0001-8933-9008]

%\ead{wjh@example.org}
%\ead[URL]{https://www.university.org}

\credit{Writing – review \& editing, Investigation, Resources}

\affiliation[2]{organization={European Commission, Joint Research Centre (JRC)}, 
                postcodesep={}, 
                city={Karlsruhe},
                postcode={D-76344},
                country={Germany}}

\author[2]{Davide Robba}
\credit{Writing – review \& editing, Investigation, Resources}

\author[2]{Luka Vlahovic}
\credit{Writing – review \& editing, Investigation, Resources}

\author[2,3]{Rudy Konings}
\credit{Investigation, Resources}

\affiliation[3]{organization={Delft University of Technology, Faculty of Applied Sciences, Department of Radiation Science and Technology}, 
                postcodesep={}, 
                city={Delft},
                postcode={2628},
                country={Netherlands}}

\author[1]{Ethan Scott}
\credit{Investigation}

\author[1,4,5]{Patrick E. Hopkins}[type=editor,
                        orcid=0000-0002-3403-743X]
\cormark[2]
\ead{peh4v@virginia.edu}

\credit{Writing – review \& editing, Supervision, Conceptualization, Investigation, Methodology, Resources, Funding acquisition, Project Administration}

\affiliation[4]{organization={Department of Materials Science and Engineering, University of Virginia}, 
                city={Charlottesville},
%               citysep={}, % Uncomment if no comma needed between city and postcode
                postcode={22904}, 
                state={VA},
                country={USA}}

\affiliation[5]{organization={Department of Physics, University of Virginia, Charlottesville, VA, USA 22904}, 
                city={Charlottesville},
%               citysep={}, % Uncomment if no comma needed between city and postcode
                postcode={22904}, 
                state={VA},
                country={USA}}

\cortext[cor1]{Corresponding author}
\cortext[cor2]{Principal corresponding author}
%\fntext[fn1]{This is the first author footnote, but is common to third author as well.}
%\fntext[fn2]{Another author footnote, this is a very long footnote and  it should be a really long footnote. But this footnote is not yet  sufficiently long enough to make two lines of footnote text.}

\begin{abstract}
Advances in next-generation hypersonic hot structures, high heat-flux fusion or fission components, and laser based additive manufacturing depend on reliable solid state thermal conductivity data at high and ultrahigh temperatures, where conventional measurements become increasingly sensitive to contact resistances, uncertain boundary conditions, and nonlinear radiative losses. Building on our initial demonstration of ultrahigh temperature steady-state temperature differential radiometry (SSTDR), we present a substantially more robust platform aimed at making high temperature thermal and radiative property measurements more routine. The method integrates lock-in infrared thermography with a spatially localized, modulated perturbation laser to form a conduction dominant differential observable. We rely on in situ hyperspectral pyrometry to provide true temperature and normal spectral emittance (absorptivity), constraining absorbed power at the laser wavelengths and the measured differential temperature response under evolving surface states. We further develop a validated 2D axisymmetric steady state heat transfer model with multiple laser heat-flux inputs and converged nonlinear radiative (and optionally convective) boundary conditions, together with an analysis workflow that reconciles the quasi-steady periodic experiment with the steady-state forward model used for fitting. Using high purity molybdenum as a benchmark, we report solid state thermal conductivity $k(T)$ from 1500 – 3000 K (to the onset of melting) with $2\sigma$ uncertainties of $7.9-11 ~\%$ enabled by comprehensive uncertainty propagation, sensitivity analysis, and bounding studies. We additionally provide normal spectral emittance $\mathrm{\varepsilon(\lambda,T)}$ of molybdenum in both solid and liquid states over 500–1000 nm. These advances establish SSTDR as an accurate, non-contact route for closing the high temperature $k(T)$ data gap while simultaneously producing much needed phase dependent radiative property data for melt adjacent and extreme heat-flux applications.

\end{abstract}

\begin{graphicalabstract}
\centering
\includegraphics[width=1\textwidth]{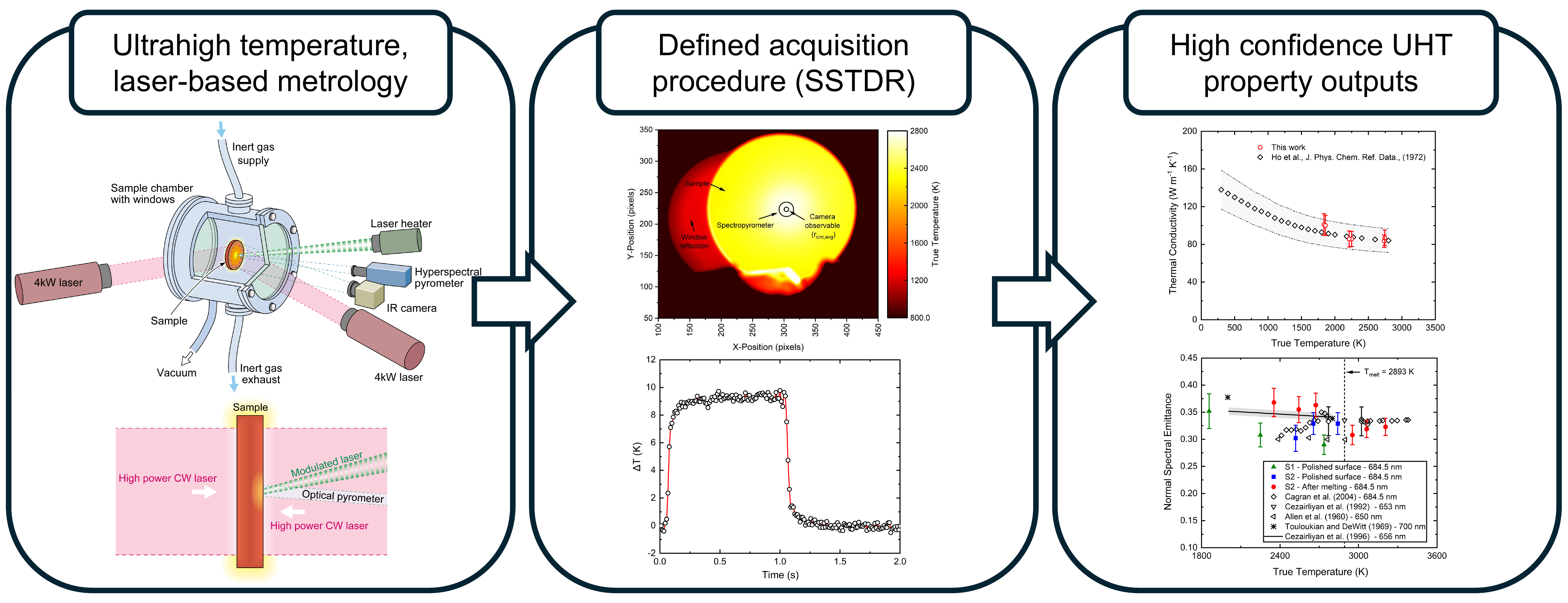}
\end{graphicalabstract}

\begin{highlights}
\item Non-contact laser based SSTDR measures solid molybdenum $k(T)$ to 3000 K
\item k(T) agrees with literature with $2\sigma$ uncertainties of $7.9$ to $11 \: \%$
\item Localized laser perturbation reduces sensitivity to boundary losses
\item In situ pyrometry constrains true temperature and laser absorptivity
\item Solid and liquid molybdenum emittance was measured from 500 to 1000 nm
\end{highlights}


\maketitle

\section{Introduction} %probably add something there that shows we do not have to assume a heat capacity

Reliable knowledge of thermophysical properties is essential for deploying materials in ultrahigh temperature environments.\cite{rapp_materials_2006,eswarappa_prameela_materials_2022,wyatt_ultra-high_2023,fahrenholtz_ultra-high_2017,baino_polymer-derived_2021,savino_arc-jet_2010,shin_advanced_2020,squire_material_2010,wuchina_designing_2004} Recent needs spanning hypersonic hot structures and space re-entry, high heat-flux fusion or fission components, and laser-based additive manufacturing have renewed demand for quantitative property data at temperatures exceeding $ \mathrm{2,000~ K}$.\cite{wuchina_uhtcs_2007,peters_materials_2024,jalali_capturing_2024,bloom_challenge_1998,de_bianchi_thermo-elastic_2021,tejado_wcu_2020,v_muller_tailored_2020,mungiguerra_qualification_2022,mungiguerra_arc-jet_2019,justin:hal-01183657} Among the most critical inputs for thermal management and energy balance modeling are thermal conductivity, emissivity (or emittance)\footnote{Emissivity is the intrinsic radiative capacity of a material. Emittance is the radiative capacity of a real surface, which includes the intrinsic factor (emissivity) in addition to extrinsic factors such as imperfections, surface roughness, porosity, grain size, etc.}, and melting behavior.\cite{zhang_advanced_2024,zhao_high-throughput_2005,saad_radiation_2023,marschall_high-enthalpy_2010,tang_ablation_2007,cedillos-barraza_investigating_2016} Metals are particularly attractive in high heat flux settings because high thermal conductivity suppresses adverse temperature gradients and enables efficient heat spreading.\cite{tenney_materials_1989,gold_refractory_1979,kasen_heat_2019,linsmeier_development_2017} In melt-adjacent processes such as laser based manufacturing, accurate high temperature thermal and optical property inputs are also required to model temperature fields, stability near developing melting, and re-solidification behavior.\cite{wang_situ_2022,khorasani_effect_2022,dobbelstein_direct_2016,chen_three-dimensional_2017,dai_thermal_2014,ye_energy_2019,cook_determining_2023,hooper_melt_2018}

Despite their importance, robust solid state thermal conductivity measurements above  $ \mathrm{2,000~ K}$ remain limited, and reported values can diverge across methods as radiative losses and boundary condition uncertainty intensify.\cite{pfeifer_limitations_2025} Historically, ultrahigh temperature thermal conductivity $k(T)$ has often been inferred directly from electrical resistivity via the Wiedemann-Franz law.\cite{pottlacher_thermal_1999,tolias_analytical_2017} While valuable, such inference depends on assumed Lorenz numbers, electronic transport models, and material state, and it does not provide an independent metrological benchmark for validating transport models in the ultrahigh temperature regime. These limitations motivate measurement approaches that remain non-contact, minimize sensitivity to uncertain radiative and parasitic losses, and provide traceable uncertainty suitable for model validation.

Building on our initial demonstration of steady-state temperature differential radiometry (SSTDR)\cite{milich_validation_2024}, here we present a substantially more robust implementation designed specifically to close this ultrahigh temperature solid $k(T)$ gap. The approach integrates lock-in infrared thermography with a spatially localized, modulated perturbation laser to form a conduction dominant differential observable, improving sensitivity to $k$ while reducing susceptibility to uncertain radial boundary conditions and nonlinear radiative losses. In situ hyperspectral pyrometry is incorporated to determine true temperature and wavelength resolved emittance (absorptivity) that constrains absorbed power and the measured differential temperature response under evolving surface states. Importantly, SSTDR is not reliant on electrical conduction, enabling direct thermal conductivity measurements in materials dominated by electronic and phononic heat transport as well as in primarily phononic conductors, thereby broadening applicability beyond resistivity based inference methods. Because SSTDR is quasi-steady, it is effectively insensitive to heat capacity and does not require $c_{p/v}$ inputs, unlike transient flash-type techniques.\cite{pfeifer_limitations_2025} Moreover, the measurement is relatively rapid (typically tens of seconds per operating point), reducing exposure time at extreme temperature and thereby mitigating microstructural evolution, oxidation/contamination, or surface-state drift that can complicate longer duration protocols.

In parallel, we develop and validate a 2D axisymmetric steady-state heat transfer model incorporating multiple laser heat flux inputs, converged nonlinear radiative (and optionally convective) boundary conditions, and an analysis workflow that reconciles the quasi steady state periodic experiment with the steady-state forward model used for fitting. Using high-purity molybdenum as a benchmark, we report solid state thermal conductivity from $\mathrm{1,500 - 3,000 ~K}$ (up to the onset of melting) with an in depth uncertainty propagation framework. While the present benchmarks span $\mathrm{1,500-3,000 ~K}$ (limited by the molybdenum melting point), the SSTDR methodology and instrumentation are scalable to substantially higher temperatures and exceeding $\mathrm{4,000 ~K}$, highlighting a pathway to extend uncertainty quantified $k(T)$ measurements further into the ultrahigh temperature regime for higher melting materials.\cite{cedillos-barraza_investigating_2016,milich_validation_2024} We additionally provide normal spectral emittance $\varepsilon(\lambda,T)$ (NSE) of molybdenum in both the solid and liquid states over 500 - 1000 nm. While this band does not by itself constitute total hemispherical emittance for direct full spectrum energy balance enclosure, it constrains wavelength specific absorptivity at the laser wavelengths and provides phase-aware spectral radiative behavior that can inform broader radiative property models. 

Finally, we compare the measured $k(T)$ trends against resistivity based expectations as an independent cross check, while emphasizing that the primary contribution of this work is an uncertainty quantified, non-contact route for generating high confidence ultrahigh temperature solid $k(T)$ benchmarks together with phase aware spectral radiative property inputs.

\section{Methodology}

\subsection{Apparatus and acquisition}\label{sec:aparatus}

The principle of steady-state temperature differential radiometry (SSTDR) is to heat a specimen to a baseline temperature of interest, apply a small perturbative heat flux $\Delta q$, and observe the resulting change in surface temperature. Thermal conductivity is extracted by fitting the measured differential temperature response to a heat transfer model. In our original implementation, baseline heating was achieved with a single front-side infrared (IR) laser and the perturbation response was measured using single-color pyrometry.\cite{milich_validation_2024} While this established feasibility at ultrahigh temperatures, it imposed three practical limitations that motivated the present upgrades: a baseline laser spot size comparable to the specimen size increased sensitivity to uncertain radial boundary conditions and to sample mounting details; single-color pyrometry required an assumed emittance at the pyrometer wavelength to convert radiance temperature into true temperature, which can be problematic under evolving surface states and for materials with unknown properties; and without lock-in detection, the perturbation temperature signal is noisy (signal noise $\mathrm{>10~\%}$ of measured maximum differential), directly degrading the precision of the measured $\Delta T$ and thus the uncertainty in extracted $k$. In addition, single source/single sided heating can complicate temperature control during approach to quasi-steady state, where transient overshoots and baseline drift may require correction. These considerations compounded and resulted in relative uncertainties in reported data of $18-20 \: \%$.

\begin{figure*}
    \centering
    \includegraphics[width=1\textwidth]{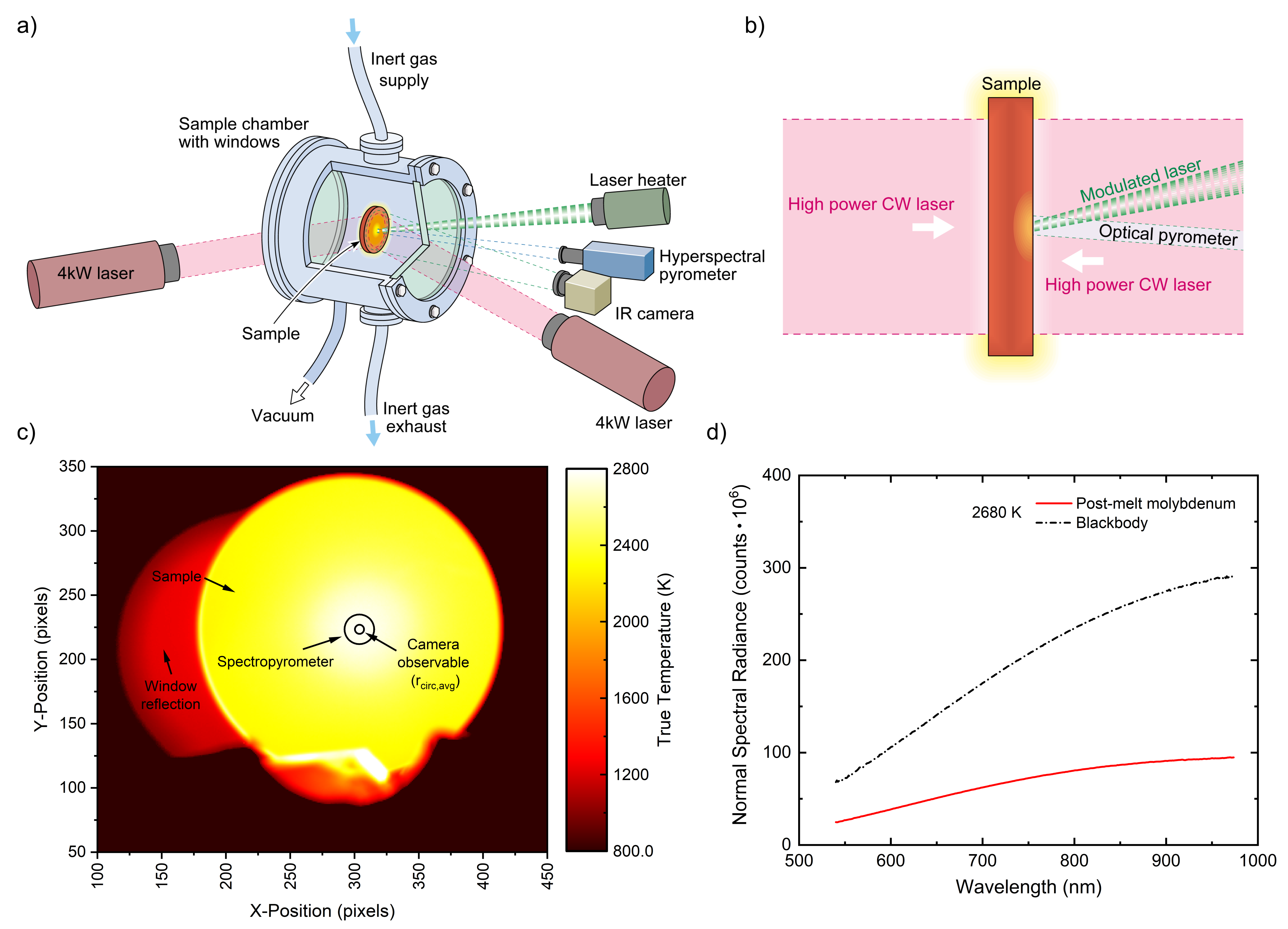} 
    \caption{\label{fig:main} a) Measurement apparatus used to conduct steady-state temperature differential radiometry (SSTDR) experiments. b) Sample disk geometry cross section depicting relative physical scale differences of localized perturbative heating, pyrometry and bulk heating. c) Infrared camera (IR) image of a disc molybdenum specimen heated to near melt (uncorrected for window reflections). Measurement field of views and locations of the hyperspectral pyrometer and the digitally formed observable with radius $r_{circ,avg}$ used to construct $\Delta T$ are overlaid. d) Normal spectral intensity measured by the hyperspectral pyrometer on post-melt molybdenum at 2680 K compared to blackbody measurements at the same temperature (corrected for optical efficiency and background). }
\end{figure*}

In the SSTDR configuration implemented here (Fig. \ref{fig:main}a), a high power continuous wave (CW) laser (Trumpf TruDisk 4001, $\lambda = 1030 ~nm$, maximum output 4 kW) established the baseline temperature, similar in principle to the original SSTDR configuration. The output laser is fiber coupled and can be directed either entirely to the front surface or split 50/50 between the front and back surfaces, each forming a uniform top hat intensity profile. The split heating configuration is primarily used to reduce through thickness temperature gradients and thermal stresses during baseline stabilization prior to perturbative heating, which is particularly advantageous for brittle ceramics (though less critical for ductile metals).\cite{vlahovic_thermal_2018}

A central experimental advance is the integration of lock-in infrared thermography with a spatially localized modulated perturbation laser to form a conduction dominant differential observable (Fig. \ref{fig:main}b). After the baseline temperature is established, a secondary low power laser (Coherent Verdi V5, $\lambda = 532 ~nm$, maximum output 5 W) is applied at the center of the baseline heated region with a Gaussian intensity profile and square-wave temporal modulation (optical chopper). A small fraction ($1 \: \%$) of the modulated beam is diverted to a photodiode whose signal is used to synchronize phase-sensitive detection in the infrared camera. The resulting lock-in measurement provides a spatially resolved, periodic temperature response with substantially improved signal-to-noise ratio (Fig. \ref{fig:main}c).

In parallel, in situ hyperspectral pyrometry is used to determine true temperature and wavelength resolved emittance (absorptivity) that constrains absorbed power and the measured differential temperature response without prior knowledge of surface properties under evolving surface states.

Front-side surface radiance is measured with a hyperspectral spectropyrometer (FAR Associates FMP2/2X) using a $\approx 2 ~mm$ diameter field of view (Fig. \ref{fig:main}a-c). The system collects emitted radiance $L_\lambda(\lambda,T)$ at 500 discrete wavelengths between 500 and 1000 nm. Measured detector signals are converted to absolute spectral radiance through full optical path calibration (including chamber window transmission, notch filters, and fiber optics) using a NIST-traceable graphite tube blackbody source (Mikron M390) at $\mathrm{2,723~ K ~(2,450 ~^{\circ}C)}$ (Fig. \ref{fig:main}d). 

The measured spectral radiance is related to the true surface temperature and normal spectral emittance through
\begin{equation}
L_\lambda(\lambda,T)=\varepsilon(\lambda,T)L_{\lambda,b}(\lambda,T),
\label{eq:spec_rad_general}
\end{equation}
where $L_{\lambda,b}$ is the blackbody spectral radiance. True temperature $T_{true}$ and spectral emittance $\varepsilon(\lambda,T)$ were determined using the multiwavelength spectropyrometry framework developed by Felice and co-workers\cite{Felice2003,felice_expert_2003,felice_temperature_1998,felice_pyrometry_2013}, in which the measured spectrum is interrogated through a large ensemble of two-wavelength temperature determinations rather than a single-color or conventional two-color estimate. When the pairwise temperatures are mutually consistent, a gray-body description is sufficient; when systematic disagreement is present, wavelength-dependent emittance is introduced into the fit.

For the linear spectral emittance model ultimately utilized here,
\begin{equation}
\varepsilon(\lambda)=a_0+a_1\lambda,
\label{eq:eps_linear_main}
\end{equation}
the corresponding pairwise ratio temperature for wavelengths $\lambda_i$ and $\lambda_j$ is
\begin{equation}
T_{ij}
=
\frac{C_2\left(\lambda_j^{-1}-\lambda_i^{-1}\right)}
{\ln\left[L_\lambda(\lambda_i)/L_\lambda(\lambda_j)\right]
-\ln\left[\dfrac{a_0+a_1\lambda_i}{a_0+a_1\lambda_j}\right]
-5\ln\left(\lambda_j/\lambda_i\right)},
\label{eq:pair_temp_linear_main}
\end{equation}
where $C_2$ is the second radiation constant. In practice, the fitting routine determines the simplest spectral emittance model that yields a statistically self-consistent temperature solution over the accepted wavelength interval; higher order emittance representations are treated analogously. More detailed description of the process is provided in previous reports.\cite{Felice2003,felice_expert_2003,felice_temperature_1998,felice_pyrometry_2013}

We note that while some previous studies have cautioned against similar approaches, in practice this methodology is viable for a vast majority of materials and does not alter the main conclusions of this work. Continuing, once an adequate emittance parameterization (gray, linear, or higher-order) is identified for the measured band (or a part of the band), the true temperature becomes well constrained, and the remaining wavelength dependent emittance across 500 - 1000 nm can then be recovered directly from the calibrated radiance spectrum. This approach is particularly valuable under evolving surface states because it does not require a priori emittance input at a single wavelength. We emphasize that these measurements provide normal spectral emittance over a finite band rather than total hemispherical emittance; accordingly, they are used primarily to constrain laser wavelength absorptivity, bound the true temperature differential and characterize phase dependent spectral radiative behavior that can inform broader radiative property models. 

For thermal imaging of the perturbation response, an infrared camera (Infratec ImageIR 8380 hp; 2-5.7 $\mu m$ spectral range) records the spatial radiance temperature field (Fig. \ref{fig:main}c). The camera employs a 640 x 512 pixel sensor and effective pixel size at the sample plane depends on the selected optics and working distance (typically $\sim$90 $ \mu m $ in the present configuration). The perturbation beam diameter is chosen to provide adequate sampling (approximately 10 pixels across the $1/e^2$ beam diameter: ~850 $\mu m$) while minimizing radial boundary effects on the induced response. Under steady conditions, thermal spreading increases the effective perturbation diameter by approximately a factor of two. Perturbation power is adjusted to maintain the small perturbation regime (<50 K rise relative to the baseline), limiting sensitivity to temperature dependent property variation. The modulation frequency is kept low (0.5 Hz in this work) to maintain quasi-steady thermal gradients ensuring insensitivity to heat capacity and facilitate the steady-state model inversion.

High  purity molybdenum (99.95$\%$, Midwest Tungsten Service) was sectioned by electrical discharge machining (EDM) into 20 mm diameter, approximately 2 mm thick disks. The baseline heating laser was set to a 5.7 mm diameter top-hat spot for all measurements. Both sample faces were polished sequentially using 600, 800, and 1000 grit SiC papers, followed by 6, 3, 1, and 0.25 $\mu m$ diamond suspensions, yielding a final surface roughness of $\sim$50 nm RMS (mirror-like surface) as measured by profilometry.

Samples were mounted vertically inside a pressure vessel using zirconia holders designed to minimize contact area and reduce conductive heat loss through the mount. The chamber incorporated uncoated $CaF_2$ windows selected for high, relatively flat transmission from the visible through the mid-IR, covering the operating bands of the lasers, hyperspectral pyrometer and the infrared camera. The vessel was evacuated to $\sim 10^{-5}$ hPa and back filled with ultrahigh purity argon (nominal 1 ppm $O_2$) to the target pressure. The argon was further purified through oxygen and moisture traps and processed through an electrochemical oxygen pump (SETNAG GEN'AIR) to reduce effective oxygen partial pressure to the $10^{-18}-10^{-20}$ bar range, as verified by an exhaust oxygen analyzer, SETNAG JOK'AIR. Once the target atmosphere was established, inlet and outlet valves were closed to isolate the system (We note that due to degassing from the walls after valve closure, the measurement $pO_2$ is likely higher). The specimen was then heated gradually to the target baseline temperature by incrementally increasing laser power, with heating/cooling rates adjusted as needed based on measurement duration and stability criteria.

Liquid state thermal conductivity can be extracted using the same SSTDR framework when melt geometry is explicitly parameterized; in the present study, molybdenum is used as a benchmark to report solid state $k(T)$ up to the onset of melting and to obtain normal spectral emittance in both solid and liquid states (a surface melt pool is formed and the sample acts as a self crucible).\cite{schonfeld_melting_2025,milich_validation_2024,manara_new_2008}

In summary, the SSTDR data acquisition proceeds as follows: 
\begin{enumerate}
  \item Load the disk sample into the chamber and if needed establish an inert, low-$pO_2$ environment.
  \item Heat the sample with the high-power laser (optionally split front/back) until quasi-steady baseline conditions are reached at the desired temperature.
  \item Use hyperspectral pyrometry to determine true surface temperature and normal spectral emittance (NSE) $\varepsilon(\lambda,T)$ within the spectropyrometer field of view. 
  \item Apply the lower power modulated perturbation laser at the center of the heated region to induce small periodic surface temperature variation. 
  \item Use lock-in infrared thermography synchronized to the modulation signal to measure the spatially resolved differential temperature response for inversion of $k$.
\end{enumerate}

\subsection{Thermal model and data analysis}

To extract thermal conductivity, $k$, the infrared (IR) camera thermograms are first converted from radiance temperature to true surface temperature to accurately determine the perturbation induced temperature response, $\Delta T$. True temperature is established using hyperspectral pyrometry (500 - 1000 nm), which provides $T_{true}$ independent of, or simultaneously with, normal spectral emittance within a defined field of view (FOV). Because the sample emittance is not measured directly in the IR camera band (2 - 5.7 $\mu$m), an effective band emittance $\varepsilon_{\perp,2-5.7}$ is determined by iterating its value until the IR-derived temperature, averaged over a circular FOV matched to the spectropyrometer footprint, equals the spectropyrometer determined $T_{true}$ (Fig. \ref{fig:main}c). 

\begin{figure*}
    \centering
    \includegraphics[width=1\textwidth]{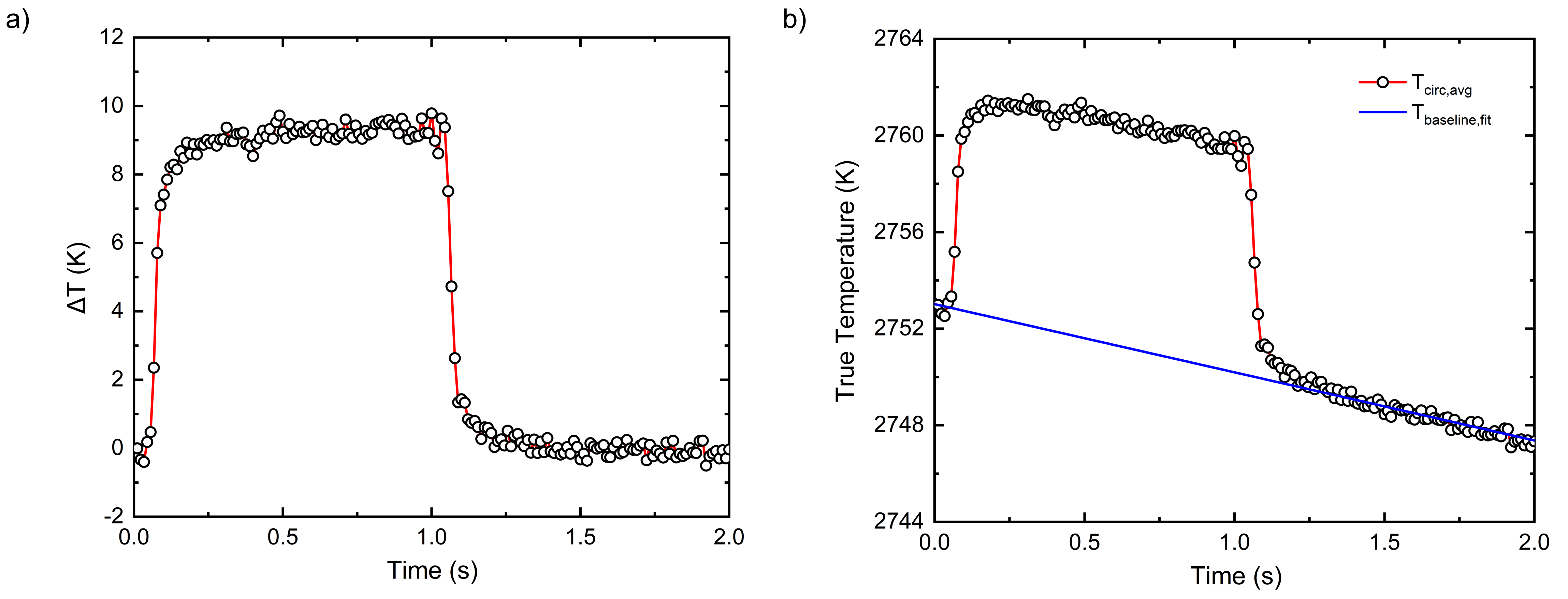} 
    \caption{\label{fig:corr} a) Measured $\Delta T(t)$ during perturbative modulation formed from the IR camera observable with radius $r_{circ,avg}$ (Fig. \ref{fig:main}c). b) Measured $T(t)$ and baseline subtraction procedure used to mitigate effect of slow bulk thermal drift induced under quasi-steady state conditions. }
\end{figure*}

After the correction, the IR camera's temperature field is circularly averaged over a radius $r_{circ,avg}$ centered on the perturbation beam to form the experimental observable $\Delta T(t)$ (Fig. \ref{fig:main}c and \ref{fig:corr}a). Sensitivity to $k$ increases as the sampling region is reduced, therefore $r_{circ,avg}$ is chosen as small as practical while maintaining acceptable noise levels. The modulation cycle temperature response is then obtained over multiple periods, and $\Delta T_{max}$ is defined as the maximum of $\Delta T(t)$ within the modulation envelope and after the perturbation response reaches a maximum quasi steady-state value. If slow baseline drift is present in the thermogram, a linear trend is removed before computing $\Delta T(t)$ so that drift does not bias the resulting $\Delta T$ used in the inversion (Fig. \ref{fig:corr}b).

The hyperspectral measurements further provide normal spectral emittance $\varepsilon(\lambda,T)$ over 500 - 1000 nm. These spectra are linearly interpolated to the 532 nm perturbation wavelength and extrapolated to the 1030 nm baseline heating wavelength to estimate wavelength-specific absorptivity used to compute absorbed powers (with window transmission corrections applied). We emphasize that the measured $\varepsilon(\lambda,T)$ is a normal spectral emittance over a finite band rather than total hemispherical emittance. Accordingly, it is used here primarily to constrain absorbed power at the laser wavelengths, the true perturbation temperature differential $\Delta T$, and to characterize phase-dependent spectral radiative behavior rather than to directly compute full spectrum radiative heat loss.

The steady-state temperature field in the axisymmetric disk specimen is computed by solving the cylindrical heat equation using radially symmetric finite-volume/finite-difference discretizations. The steady-state heat equation is solved with nonlinear radiative exchange implemented through Robin boundary conditions (i.e., boundary conditions where heat flux depends on both temperature and its gradient). Radiative losses are treated via a converged nonlinear boundary formulation in which the Stefan-Boltzmann term is linearized during iteration using a secant form that is exact at convergence.\cite{incropera_fundamentals_2013} Optional convection terms can be included through an effective heat-transfer coefficient, although under the temperature conditions in this work, radiation dominates and convection is shown to be negligible (Section \ref{sec:results}.).

In the model top and bottom, in reality (Fig. \ref{fig:main}a) front and back surfaces, receive spatially resolved absorbed heat fluxes representing the baseline heaters and the small perturbation beam. The primary 1030 nm heater is modeled using a super-Gaussian profile (order m = 8) that closely approximates a tophat profile while improving numerical convergence. The perturbation heater is represented by a Gaussian profile centered at $r = 0$. Each heater profile is normalized so that its surface integral equals the absorbed power.

A representative top surface boundary condition at $z = 0$ is shown below

\begin{subequations}
\label{eq:boundary} 
    \begin{align}
    \begin{split}
        -k\frac{\partial T}{\partial z}\Big\vert_{z=0} = \underbrace{\varepsilon_T\sigma(T^4-T_{\mathrm{amb}}^4)}_{\text{radiation}} &+ \underbrace{h_{\mathrm{conv,top}}(T-T_{\mathrm{amb}})}_{\text{convection}} \\
        &- \underbrace{q_{\mathrm{heater,top}}(r)}_{\text{absorbed heater}}, \label{appa}
    \end{split}
    \end{align}
    \\
    \begin{align}
    \begin{split}
    \varepsilon_T\sigma(T^4-T_{\mathrm{amb}}^4) \approx \\ &\underbrace{\varepsilon_T\sigma(T+T_{\mathrm{amb}})(T^2+T_{\mathrm{amb}}^2)}_{h_{\mathrm{rad}}(T)}(T-T_{\mathrm{amb}}), \label{appb}
    \end{split}
    \end{align}
    \\
    \begin{align}
    \begin{split}
    h_{\mathrm{equivalent}}(T) = h_{\mathrm{rad}} + h_{\mathrm{conv,top}}, \label{appb}
    \end{split}
    \end{align}
    \\
    \begin{align}
    \begin{split}
   -k\frac{\partial T}{\partial z}\Big\vert_{z=0}  = h_{\mathrm{equivalent}}(T-T_{\mathrm{amb}}) - q_{\mathrm{heater,top}}(r), \label{appb}
    \end{split}
    \end{align}

\end{subequations}

where $\varepsilon_T$ is the total hemispherical emittance used for the radiative exchange term, $\sigma$ is the Stefan-Boltzmann constant, $k$ is the thermal conductivity and $q_{heater,top}(r)$ is the absorbed heat-flux distribution. Here $T$ denotes the surface temperature as a function of radius and $T_{amb}$ is the ambient temperature. During each Picard iteration, the nonlinear radiation term is recast into an equivalent form using a temperature dependent radiative heat transfer coefficient $h_{rad}(T)$ (secant linearization, Eq. \ref{eq:boundary}b), optionally augmented by a convective contribution $h_{conv}$ to form ($h_{equivalent}$, Eq. \ref{eq:boundary}c). The coefficients are evaluated using the temperature field from the previous iteration, so each Picard step solves a linear system while the full nonlinear radiative behavior is recovered at convergence.

Numerically, a sparse linear system is assembled at each Picard iteration to treat the nonlinear boundary terms. Axis regularity is enforced via symmetry at $r=0$, and Robin conditions with additive heater terms are applied on the outer radius, top, and bottom surfaces. Linear systems are solved with a direct sparse solver (SciPy's spsolve), and convergence is stabilized using under relaxation and gently ramped sources.

Operationally, the baseline steady-state field $T_{base}(r,z)$ is computed using only the baseline heaters (Fig. \ref{fig:bridge}a). The perturbation beam is then added to obtain as second solution $T_{perturb}(r,z)$, and the model predicted differential field is defined as $\Delta T_{sim} = T_{perturb}(r,z) - T_{base}(r,z)$. The simulated observable is taken as the circular average of $\Delta T_{sim}(r,0)$ over the same $r_{circ,avg}$ used experimentally (Fig. \ref{fig:bridge}b).

To reconcile the quasi steady periodic experiment with a purely steady state forward model, we apply an "edge correction" in which the rim value $\Delta T_{sim}(r=R)$ is subtracted prior to averaging. This suppresses residual offsets associated with slow baseline drift and weak global temperature changes not captured during the modulation period, improving correspondence between the steady-state model and the quasi-steady observable (see Section \ref{sec:bridging}.). Finally, an outer inverse loop updates the trial thermal conductivity $k$ until the surface averaged model output matches the experimental target ($|\Delta T_{sim} - \Delta T_{meas}| \leq \Delta T_{tol}) $.

\subsection{Bridging quasi-steady state measurements with steady-state modeling \label{sec:bridging}}

\begin{figure}
\centering
\includegraphics[width=1\linewidth]{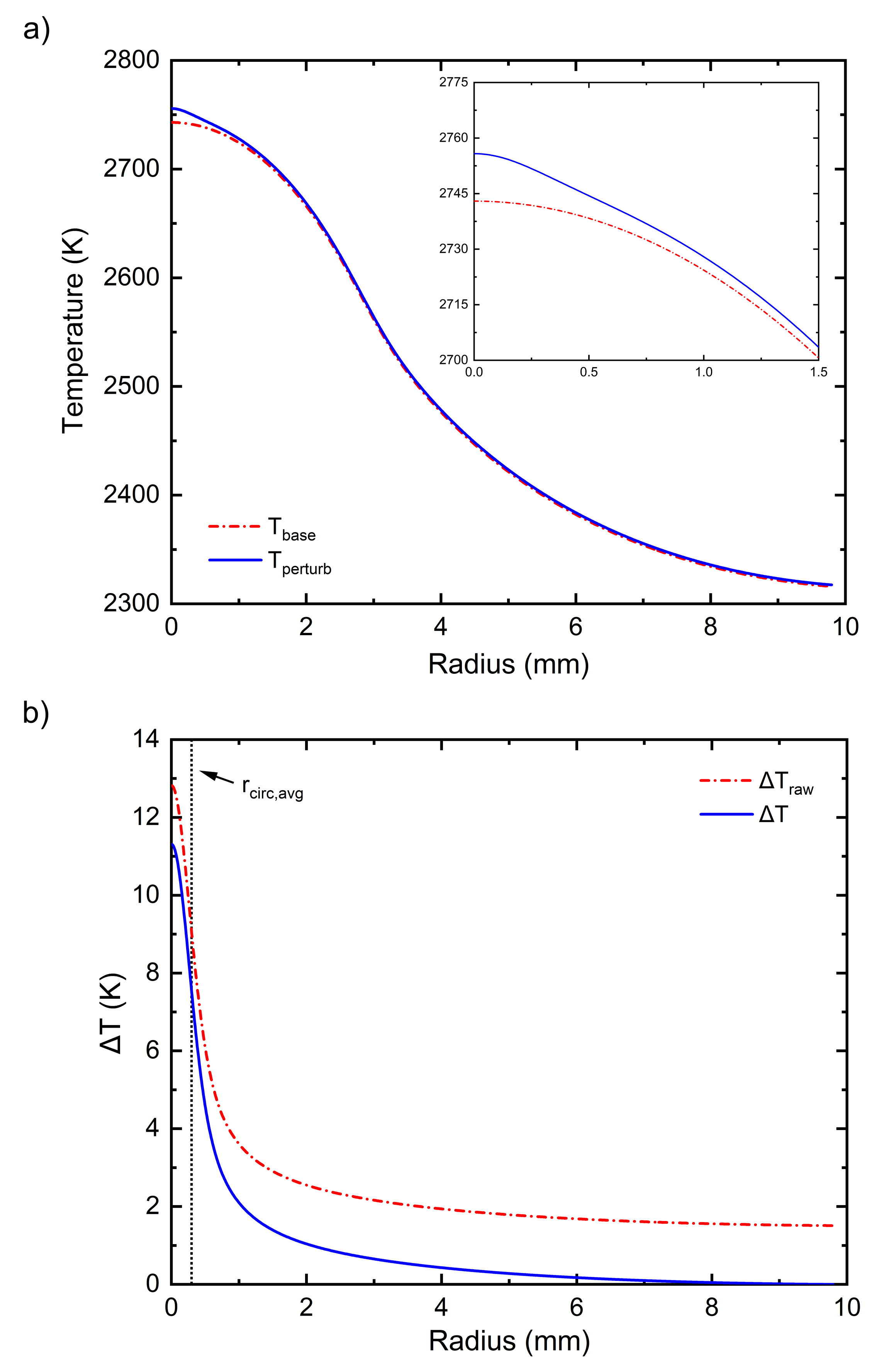}
\caption{\label{fig:bridge} a) The top surface radial profiles $T(r)$ of separately computed baseline and perturbed temperature fields. b) Edge correction conducted to separate the spatially decaying perturbation field from the uniform field offset induced by the finite, weakly lossy sample. }
\end{figure}

In the experiment, the perturbation beam is modulated, and the measured response is periodic, quasi-steady temperature rise. In the forward model, we instead solve a pair of purely steady-state problems: a baseline solution driven by the primary heaters, and a second solution with an additional (DC) perturbation source applied. Because the specimen is finite and weakly lossy, the additional source can introduce a small global temperature uplift (a near-uniform offset) that rides on top of the spatially decaying perturbation field. In practice, this appears as a nonzero rim offset in the simulated differential temperature profile $\Delta T(r)$ where the curve decays radially but does not return exactly to zero at the disk edge (Fig. \ref{fig:bridge}b).

The rim offset is a modeling/analysis complication of relating quasi-steady experimentation with a pure steady-state solution, rather than a signal of interest. The SSTDR observable that constrains the thermal parameters (e.g., $k$ and radiative-loss parameters) is the spatially varying, localized response produced by the perturbation and not a uniform background associated with slow DC accumulation. If left uncorrected, the rim offset artificially increases the circularly averaged $\Delta T$ used for fitting and can bias the inferred thermal conductivity high (the inversion must increase $k$ to reduce the model $\Delta T$ to the fixed experimental target).

To remove this uniform component, we define the top surface temperature profiles with the perturbation flux off and on as $T_{base}(r)$ and $T_{perturb}(r)$, and the raw differential as

\begin{equation}
\Delta T_{raw}(r) = T_{perturb}(r) - T_{base}(r).
\end{equation}

If the perturbation contributed only localized heating, $\Delta T_{raw}(r)$ would decay to zero as $r$ approaches the outer edge. When a nonzero rim value is present, we estimate an edge based offset

\begin{equation}
\Delta T_{raw,edge} = \Delta T_{raw}(r_{edge})
\end{equation}

taken as the differential temperature at the disk edge. The offset corrected profile is then

\begin{equation}
\Delta T(r) = \Delta T_{raw}(r) - \Delta T_{raw,edge}.
\end{equation}

All subsequent circular averages are taken on the corrected $\Delta T(r)$. For direct experimental comparison, the circular disk average with radius $r_{circ,avg}$ (same radius used to compute $\Delta T_{meas}$) is 

\begin{equation}
\Delta T_{sim} = \frac{2}{r_{circ,avg}^2} \int_{0}^{r_{circ,avg}} \Delta T(r) r d r.
\end{equation}

Physically, in the model and under complete steady-state conditions, although the perturbation beam is much lower power than the baseline heater, it still adds a finite DC load. In a finite body with finite radiative/convective loss, this can produce a near uniform uplift that is superimposed on the localized response (Fig \ref{fig:bridge}b). Removing the uniform component ensures the fit is driven by the amplitude and shape of the decaying perturbation mode. The rim offset diminishes at higher absolute temperature because radiative losses increase rapidly with $T$, so the system sheds DC heat more effectively, consistent with the smaller rim offsets observed at elevated $T$. For example, the simulated $\Delta T_{raw,edge}$ is $\mathrm{5.49~K}$ at $\mathrm{1,860 ~K}$ but decreases to $\mathrm{1.51~K}$ at $\mathrm{2,750~K}$.

As a screening criterion, we require that $\Delta T_{raw}(r)$ approach a constant value as $r$ approaches the disk edge (i.e., the remaining variation in the outer region is small compared with the central perturbation amplitude). Data sets that fail this criterion indicate insufficient experimental geometry for the edge correction to be effective and should be excluded from analysis. No such instances occurred in this study due to chosen experimental geometries.

\subsection{Uncertainty and sensitivity analysis}\label{sec:unc-sens}

We quantify the uncertainty of the fitted thermal conductivity $k$ by propagating uncertainties from the measurement and model inputs through the full analysis workflow. Each parameter is independently perturbed by its two-sigma $2\sigma$ uncertainty, the complete numerical fitting procedure is rerun to obtain a new best-fit $k$, and the resulting per-parameter variation in $k$ are combined in root sum square (RSS) to yield the total uncertainty. 

The uncertainty in true surface temperature $T_{true}$ (obtained from the hyperspectral pyrometer) is evaluated by RSS combination of three contributions: the regression/fitting uncertainty associated with converting measured spectral radiance to true temperature, instrument repeatability, and calibration uncertainty associated with the blackbody reference. This temperature uncertainty directly propagates to the perturbation induced surface temperature rise $\Delta T$, which is measured by the IR camera over the same circular field of view (FOV) used for spectropyrometry.

In addition to the temperature fit terms above, $\Delta T$ carries a noise component associated with the imaging chain and the finite magnitude of the induced signal. In the present work, the modulated source is a low-power 532 nm laser. For materials with high reflectivity and low absorptivity at this photon energy coupled with high thermal conductivity, the resulting $\Delta T$ can be small (compared to the noise floor), reducing signal to noise ratio. Increasing the modulated beam power would near proportionally increase $\Delta T$ and reduce the fractional contribution of imaging noise to the overall uncertainty budget, provided the perturbation remains in the small signal regime.

Absorptivity used to set absorbed power at the high power (1030 nm) and perturbative (532 nm) wavelengths is derived from measured emittance using Kirchhoff's law under local thermal equilibrium. To propagate the true temperature uncertainties into these absorptivities, we use the Wien approximation to Planck's law.\cite{allen_spectral_1960} Wavelength dependent radiance temperatures are computed, the true temperature is perturbed by its one-sigma value, emittance is reevaluated at each wavelength, and the corresponding absorptivity is updated. The absorptivity is then brought to its $2\sigma$ uncertainty value. These perturbed absorptivities feed the boundary heat flux terms in the thermal model for both the high-power and perturbative heaters. (Note: the radiative boundary-condition total hemispherical emittance $\varepsilon_T$ used in the heat transfer model is treated separately and taken from literature as described below.)

Additional inputs that influence the best fit thermal conductivity $k$ include:

\begin{itemize}
  \item Beam diameters: The primary heater diameters $D_{\mathrm{heater}}$ are determined from manufacturer at focus, the diameter of the perturbative laser heater $D_{\mathrm{perturb,1/e^2}}$ is measured at the focal plane with a beam profiler (uncertainty $\approx$  $1.5\: \%$)
  \item Total hemispherical emittance: $\varepsilon_{\mathrm{T}}$ of the surface for radiative boundary conditions (from literature)\cite{touloukian1969thermophysical, matsumoto_hemispherical_1999}
  \item Pixel size: The effective IR camera pixel size $L_{\mathrm{pixel}}$, used to convert the camera image to physical length scales and to set the circular disk-averaging radius $r_{circ,avg}$ for computing $\Delta T$, is determined from known specimen diameter measured in thermogram. 
  \item Specimen thickness: $H$ measured with a micrometer and varies slightly with temperature (on the order of 1 \% expansion and shown to have negligible effect).\cite{edwards_high_1951} 
\end{itemize}

For each uncertainty input to thermal conductivity $\sigma_{k,i}$ in the set

\begin{equation}
\begin{aligned}
 i = \{ \Delta T,\; \varepsilon_{\mathrm{T}},\; H,\; L_{\mathrm{pixel}}\;, D_{\mathrm{perturb,1/e^2}},\; D_{\mathrm{heater}},\; \\  A_{\mathrm{heater}},\; A_{\mathrm{perturb}} \}
\end{aligned}
\end{equation}

we apply multiplicative $\pm \; 2 \sigma $ perturbation (e.g., an absorptivity $A = 0.303$ with $7 \: \%$ uncertainty is perturbed to $A^{\pm} = 0.303(1 \pm 0.07)$, rerun the full steady-state forward model and $k$-fitting loop, and record the resulting best-fit $k^-$ and $k^+$. The per-parameter contribution to the uncertainty in $k$ is then $\sigma_{k,i} = |k_i^+-k_i^-|/2$. Finally, assuming independence across parameters, the total uncertainty is obtained by root sum square:

\begin{equation}
\sigma_k = \sqrt{\sum_i \sigma_{k,i}^2}.
\end{equation}

In addition to determining and propagating measurement uncertainties, we quantify how the perturbative flux response constrains both the bulk thermal conductivity $k$ and the radiative loss parameters. We perform local sensitivity analysis using the axisymmetric forward model. For each choice of $D_{\mathrm{perturbation,1/e^2}}$ the model computes a baseline steady-state temperature field $T_{base}(r,z)$ under the large laser heaters, and a perturbed field when the small perturbation beam is applied, thus predicting $\Delta T(r,z)$. The experimentally observed $\Delta T$ is taken as the steady-state circularly averaged surface temperature rise, as described previously.

For a given operating point, we define dimensionless sensitivities of $\Delta T$ with respect to each parameter $p \in \{k, \varepsilon_{\mathrm{T,surface}}, \varepsilon_{\mathrm{T,side}}, h_{\mathrm{conv,side}}\}$ as

\begin{equation}
S_p = \frac{\partial \ln{\Delta T}}{\partial \ln{p}}.\end{equation}

Physically, $S_p$ measures the fractional change in $\Delta T$ induced by a small fractional perturbation of $p$. A $1 \: \%$ change in $p$ produces approximately $S_p \: \%$ change in $\Delta T$. We treat the "surface emittance" degree of freedom $\varepsilon_{\mathrm{T,surface}}$ as a common value applied to both the front and back faces of the disk, while the sidewall emittance $\varepsilon_{\mathrm{T,side}}$ is allowed to vary independently to isolate the impact of radial radiative losses. The sidewall heat-loss strength is further parameterized by an effective convective coefficient $h_{\mathrm{conv,side}}$, which accounts for both true gas-phase convection and small conductive leaks into the sample holder.

Sensitivities are evaluated numerically using centered finite differences around a nominal parameter set $(k_0$\\$, \varepsilon_{\mathrm{T,surface,0}}, \varepsilon_{\mathrm{T,side,0}}, h_{\mathrm{conv,side,0}})$. For $k$ and $h_{\mathrm{conv,side}}$, we apply symmetric relative perturbations $p_\pm = p_0(1 \pm\delta)$ with $\delta = 0.03$ for $k$ and $\delta = 0.03$ for $h_{\mathrm{conv,side}}$. For emissivities we apply symmetric absolute perturbations $\varepsilon_{\mathrm{T,\pm}} = \varepsilon_{\mathrm{T,0}} \pm \delta$, for $\delta = 0.03$ which ensures a finite step size even at moderate $\varepsilon_{\mathrm{T}}$. The derivative $\partial \Delta T/\partial p$ is then approximated as

\begin{equation}
\frac{\partial \Delta T}{\partial p} \approx \frac{\Delta T(p_+) - \Delta T(p_-)}{(p_+ - p_-)}\end{equation}

and substituted into the expression for $S_p$ above. Because the perturbation beam size is adjustable experimentally, we examine how the sensitivities vary with radius. For each radius in a prescribed list, we recompute the forward solution and the corresponding sensitivity set. The resulting uncertainties and sensitivity curves are reported in Section \ref{sec:results} to access parameter leverage (e.g., conduction dominant conditions) and to bound the influence of uncertain sidewall or holder losses.

\section{Results and Discussion\label{sec:results}}

\begin{figure*}
    \centering
    \includegraphics[width=1\textwidth]{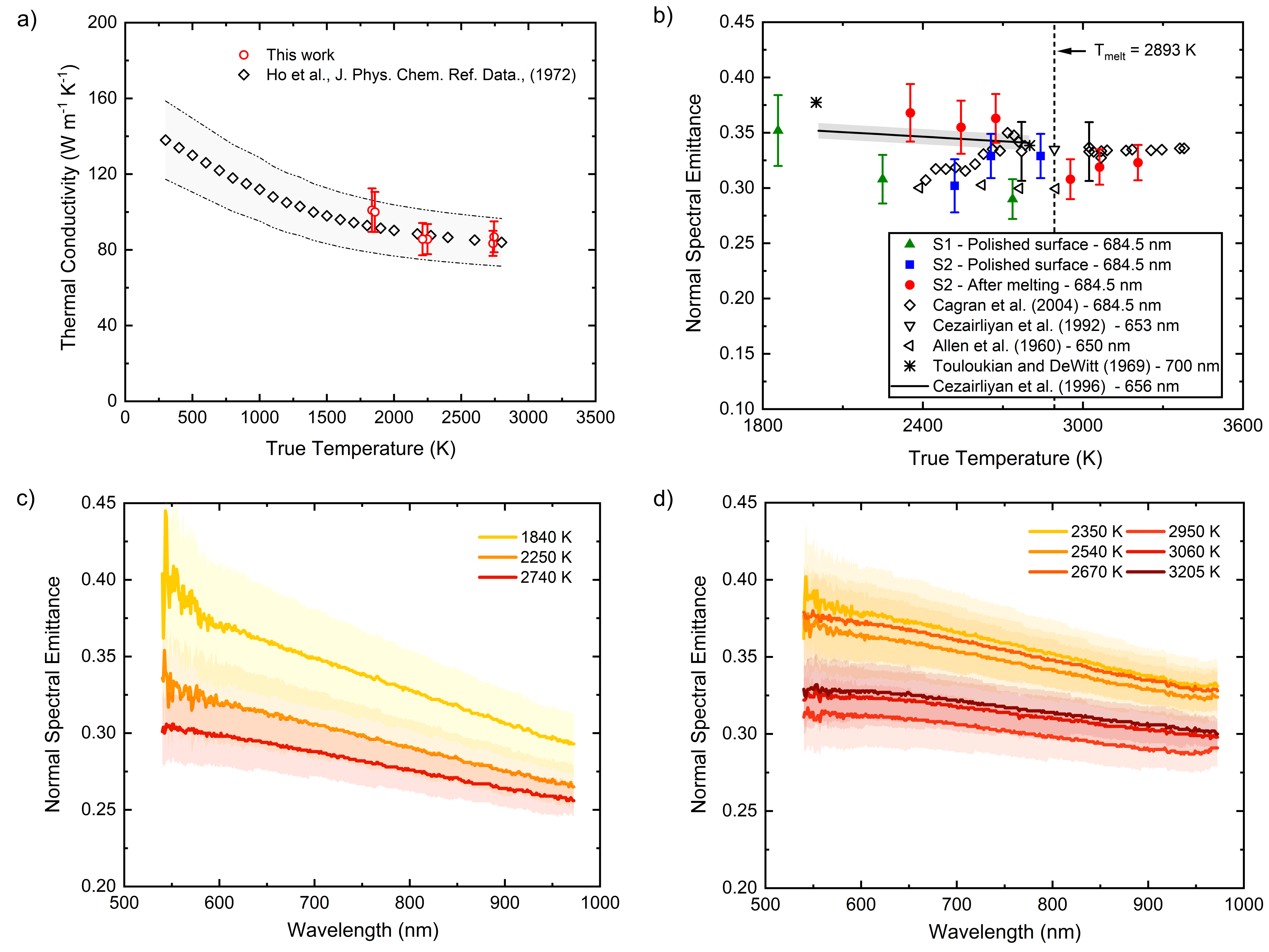} 
    \caption{\label{fig:results} a) SSTDR measured thermal conductivities of solid molybdenum compared to accepted literature values \cite{ho_thermal_1972}. b) Normal spectral emittance $\varepsilon(\lambda,T)$ of molybdenum in the solid and liquid states at 684.5 nm compared to literature values \cite{cagran_spectral_2005,cezairliyan_radiance_1992, allen_spectral_1960, touloukian1969thermophysical,cezairliyan_simultaneous_1996}. S1 and S2 refer to different samples. The melting temperature denoted by the vertical dashed line is from previously reported literature.\cite{cezairliyan_radiance_1992} c) $\varepsilon(\lambda,T)$ of polished (mirror-like) solid molybdenum S1 across the acquisition range. d) $\varepsilon(\lambda,T)$ of liquid and post-melt solid molybdenum S2 surfaces. a-d) $2\sigma$ uncertainties in reported temperature range, from $\mathrm{16 - 22 \:K}$}
\end{figure*}

The SSTDR measured thermal conductivities $k(T)$ of high purity molybdenum in the solid state is shown in Fig. \ref{fig:results}a. Across the investigated temperature range, the present measurements agree with accepted literature values within uncertainty.\cite{ho_thermal_1972,cagran_normal_2004} Because much of the available high temperature molybdenum literature infers $k(T)$ indirectly from electrical resistivity using the Wiedemann-Franz law, these data provide an independent, non-contact benchmark for solid state transport at ultrahigh temperatures, consistent with resistivity based measurements expectations and prior high temperature measurements; this conclusion is consistent with our prior report on tungsten.\cite{milich_validation_2024}

The uncertainty bounds (reported as $2\sigma$) are small relative to the spread in the published datasets, emphasizing the strong sensitivity of SSTDR to $k$ under these conditions. For molybdenum, the fractional uncertainty in the extracted thermal conductivity is $\mathrm{7.9 - 11 ~\%}$ across $\mathrm{1,800-2,800 ~K}$. This represents a substantial improvement over the $\mathrm{\sim 20~\%}$ uncertainties reported in our prior single laser heater SSTDR implementation.\cite{milich_validation_2024} Table \ref{tab:table_2sigma} breaks down per-parameter contributions from the forward model perturbations described previously. The total uncertainty is dominated by the measured differential temperature signal $\Delta T$ and the absorbed power of the perturbative heater, while other inputs are negligible.

Several aspects of the present methodology drive this improvement. First, lock-in thermography suppresses measurement noise and improves repeatability of the perturbation response used to form $\Delta T$. Second, wavelength specific absorptivity at the laser wavelengths is constrained in situ from hyperspectral measurements rather than assumed from literature, directly reducing uncertainty in absorbed power. Third, the spatially localized perturbation laser generates a confined response that reduces sensitivity of $\Delta T$ to global radial boundary conditions (e.g., distributed radiative losses over the full specimen surface and potential holder losses). Finally, a validated two-dimensional axisymmetric model with converged nonlinear radiative boundary conditions reduces modeling error in the inversion and supports rigorous uncertainty propagation.

The resulting normal spectral emittance (NSE), $\varepsilon(\lambda,T)$, of high-purity molybdenum in the solid and liquid states over 500-1000 nm is shown in Fig. \ref{fig:results}b-d. Because these measurements are performed simultaneously with hyperspectral temperature extraction, they provide phase-resolved radiative property data under the same thermal conditions used for the $k(T)$ inversion. We emphasize that the reported $\varepsilon(\lambda,T)$ corresponds to normal spectral emittance over a finite wavelength band and should not be interpreted as total hemispherical emittance.

Figure \ref{fig:results}b compares $\varepsilon(\lambda,T)$ at 684.5 nm for three surface states and across two samples (S1 and S2). The initial polished solid surfaces ($\sim$50 nm RMS starting roughness), the molten surface, and the post-melt resolidified surface. Over the solid state interval from approximately $\mathrm{1840~K}$ to $\mathrm{2800~K}$, the NSE at 684.5 nm decreases gradually with increasing temperature for S1, whereas S2 remains approximately constant, with at most a weak positive temperature dependence that falls within the experimental uncertainty. At the onset of melting for S2, we observe a decrease in NSE from 0.329 at $2,840$ K to 0.308 at $2,950$ K, with partially overlapping uncertainties. This trend is consistent with prior reports showing that the visible band emittance of molybdenum decreases upon melting.\cite{cagran_spectral_2005,urban_measuring_2024}

The post melt resolidified S2 surface also exhibits a modestly higher NSE at comparable temperature, with a value of 0.363 at $2,670$ K relative to 0.329 for the initially polished surface at $2,655$ K. Considering that NSE is highly sensitive to surface condition and, in particular, roughness, this increase in NSE suggests a modified surface morphology of the resolidified material relative to the as polished state. Indeed, optical microscopy of the resolidified region revealed microcracking near the center of the melt pool as well as distinct resolidification wavefronts. Similar, though smaller, differences are also observed between the initially polished S1 and S2 surfaces. Although most of these variations remain within the uncertainty bounds, minor differences may arise from small changes in polishing induced roughness or from the formation of slight surface oxidation in one experiment relative to the other. The influence of surface roughness on optical properties is well established. For example, He et al. showed that for molybdenum at approximately 1300 K, the emittance at 1000 nm increased from about 0.45 for a roughness average of 42 nm to about 0.60 for 103 nm roughness.\cite{he_influence_2024} Cezairliyan (1973) and Cezairliyan et al. (1992) attributed deviations in the radiance temperature of solid niobium and other metals at their melting points to changes in surface roughness. \cite{cezairliyan_1973,cezairliyan_radiance_1992} When compared with available literature at the same, or closely matching, wavelengths, the present measurements show excellent agreement in both the solid and liquid states despite differences in measurement technique and sample surface conditions.\cite{cagran_spectral_2005,cezairliyan_radiance_1992,urban_measuring_2024,cezairliyan_simultaneous_1996,touloukian1969thermophysical,allen_spectral_1960}

Figures \ref{fig:results}c and \ref{fig:results}d show $\varepsilon(\lambda,T)$ as a function of wavelength for the S1 polished solid surface, the S2 liquid surface, and the S2 post-melt resolidified surface. Across the 500 -1000 nm band, the spectra exhibit an approximately linear decrease with increasing wavelength, consistent with the expected transition from stronger interband contributions at shorter wavelengths toward increasingly free electron dominated behavior at longer wavelengths in refractory metals. The reported $2\sigma$ uncertainties in NSE range from $\mathrm{3.3-12~\%}$ depending on wavelength and temperature, and are dominated by uncertainty in the extracted true surface temperature used in the radiometric inversion.

\begin{table}
\caption{\label{tab:table_2sigma} Per-parameter contributions to uncertainty for fitted thermal conductivity of single experiments. Uncertainty in reported true temperature is $\mathrm{16-22 ~K}$.}
\begin{tabular*}{\tblwidth}{@{} CCCCC@{} }
% Changed {ccddd} to {ccccc} so all 5 columns are simple centered text
%\begin{tabular}{ccccc}
\midrule
Experiment & $i$ & $2\sigma_i$ & $2\sigma_{k,i}$ & $2\sigma_{k,i}$\\
 &  & $(\%)$ &$ (W \:m^{-1} \:K^{-1})$ & $(\%)$ \\
\hline
\\[-4pt]
Mo:1 (2,736 K) &  &  &  & \\
$83.43 \;(W \:m^{-1} \:K^{-1})$& & & &  \\
 & $\Delta T$ & 3.72 & 3.00 & 3.60 \\
 & $\varepsilon_{\mathrm{T}}$ & 13.5 & 0.07 & 0.08 \\
 & $r_{circ,avg}$ & 1.36 & 0.41 & 0.49 \\
 & $H$ & 1.4 & 0.09 & 0.11 \\
 & $D_{\mathrm{perturb}}$ & 1.5 & 0.78 & 0.93 \\
 & $D_{\mathrm{heater}}$ & 7 & 0.29 & 0.35 \\
 & $A_{\mathrm{perturb}}$ & 7.18 & 5.82 & 6.98 \\
 & $A_{\mathrm{heater}}$ & 4.02 & 0.04 & 0.05 \\
 & & & &  \\
Total (RSS): & & & 6.61 & 7.92 \\

%\midrule

\\[-4pt]
Mo:6 (1,836 K) &  &  &  & \\
$100.98 \;(W \:m^{-1} \:K^{-1})$& & & &  \\
 & $\Delta T$ & 2.94 & 2.97 & 2.94 \\
 & $\varepsilon_T$ & 15 & 0.26 & 0.26 \\
 & $r_{circ,avg}$ & 1.36 & 0.49 & 0.49 \\
 & $H$ & 1.4 & 0.08 & 0.08 \\
 & $D_{\mathrm{perturb}}$ & 1.5 & 0.94 & 0.93 \\
 & $D_{\mathrm{heater}}$ & 7 & 0.09 & 0.09 \\
 & $A_{\mathrm{perturb}}$ & 10.96 & 11.02 & 10.91 \\
 & $A_{\mathrm{heater}}$ & 7.16 & 0.46 & 0.46 \\
 & & & &  \\
Total (RSS): & & & 11.48 & 11.36 \\
\bottomrule
%\end{tabular}
\end{tabular*}
\end{table}

Although radiative losses are small for the present molybdenum conditions, this will not be true for all SSTDR implementations. Larger perturbation footprints, thicker specimens, lower thermal conductivity materials, or operation at higher temperatures will increase the radiative contribution to the perturbed temperature field. Under those conditions, radiation can measurably influence the differential response used to fit $k$, and accurate radiative boundary conditions become correspondingly more important. While this is the case, the same methodological approach and sensitivity and bounding framework defined here, can be similarly applied, which is a primary goal of this work.

Looking forward, the most direct path to further reducing SSTDR uncertainty is to mitigate measurement noise in the differential temperature signal $\Delta T$. In the present uncertainty budget, approximately $95 ~\%$ of the $\Delta T$ uncertainty is attributable to imaging noise (with the caveat that at high temperature, and depending on detector band, $\Delta T$ can become more sensitive to emittance variation within the measurement band). This noise sensitivity is most pronounced when $\Delta T$ is small, as occurs for highly conductive materials with low absorptivity at the perturbation wavelength. Operating conditions that increase the perturbation induced temperature rise and target $\Delta T$ in the $\sim 20-40~K$ range at the temperatures studied in this work would reduce the fractional noise contribution. The remaining dominant contributor is the absorbed power of the perturbative heater, which is constrained here by in situ wavelength specific absorptivity, leaving comparatively limited headroom without changing the optical configuration or calibration strategy.

\begin{figure}
\centering
\includegraphics[width=1\linewidth]{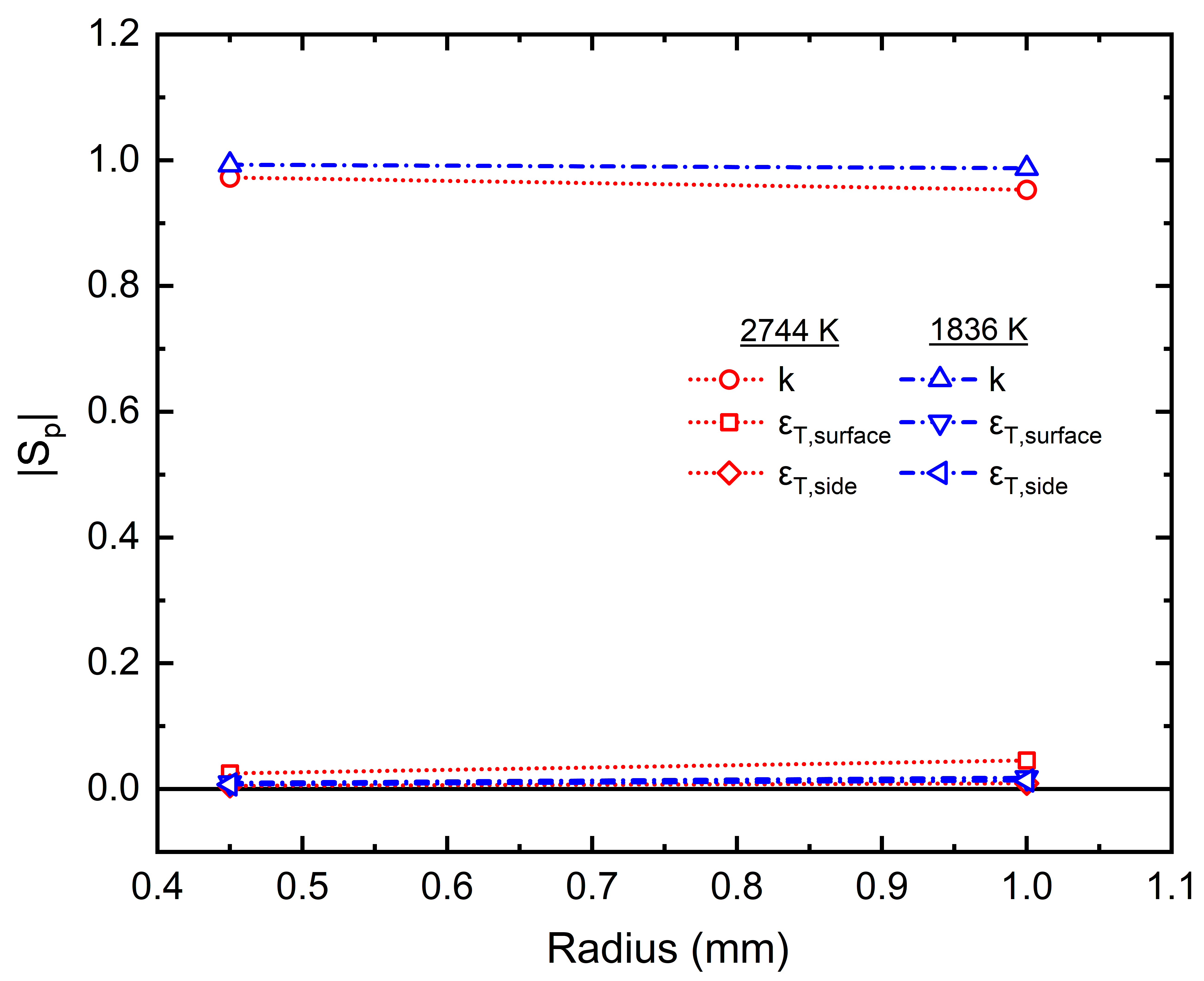}
\caption{\label{fig:sensitivity} Sensitivity of parameter $p$ to the measured SSTDR observable $\Delta T$ as a function of perturbation laser incident spotsize.}
\end{figure}

In addition to per-parameter perturbation studies used for uncertainty propagation, we performed a sensitivity analysis to quantify how $\Delta T$ responds to variations in bulk thermal conductivity and emittance for the experimental geometries considered here (as discussed in Sec. \ref{sec:unc-sens}). The analysis was carried out across multiple temperatures and for several perturbation spotsizes consistent with current power levels (larger spots are accessible with higher power perturbation lasers or when testing lower $k$ materials, where signal to noise constraints are less limiting). Figure \ref{fig:sensitivity} shows that $\Delta T$ remains strongly conduction dominated for these specimens due to the sample geometry and spatially localized perturbation. The magnitude of the thermal conductivity sensitivity is approximately unity, $|S_k| \approx 1$, at both moderate and ultrahigh temperatures and across the spotsize range considered. In contrast, emittance sensitivities are near zero ($|S_{\varepsilon_T}| \approx 0$), increasing only slightly with increasing temperature and perturbation spotsize. This behavior is consistent with conductive heat spreading away from the perturbation region dominating the local heat flux and $\Delta T$ with minor effects from radiative exchange. For lower thermal conductivity materials or larger perturbation footprints (when compared to sample radius), the relative importance of radiation is expected to increase and the corresponding $|S_{\varepsilon_T}|$ values should rise accordingly. While the differential observable is weakly emittance dependent, a reasonable estimate of total emittance remains important for reproducing the correct baseline temperature field in the forward model prior to fitting the differential response. 

We also assessed whether gas convection from sample surfaces or heat loss through the sample holder contact could bias the reported thermal conductivity values. Although the boundary conditions framework allows convection, the nominal results assume zero convective flux. To bound any possible effect, we repeated the analysis for the lowest and highest temperature solid molybdenum cases by adding a free convection term with a conservative upper bound coefficient  $h_{\mathrm{conv}} =$ 25 W m$^{-2}$ K$^{-1}$, consistent with standard textbook ranges for gases ($ \approx 2-25$ W m$^{-2}$ K$^{-1}$).\cite{incropera_fundamentals_2013} This coefficient was applied to all exposed boundaries (front face, back face, and radial surface), and the data were refit using the nominal procedure. Including convection produced a negligible change in the extracted thermal conductivity as the fitted $k$ increased by $0.08~\%$ at 1,836 K and by $0.01~\%$ at 2,744 K, which is within numerical tolerance. This indicates that, for the present geometry and operating conditions, the measured $\Delta T$ is effectively insensitive to convective transfer.

Similarly, we evaluated whether conductive heat loss through the localized, insulating sample holder could bias the fitted thermal conductivity. Because the true holder contact is localized and non-axisymmetric, it cannot be represented directly in the 2D axisymmetric model. Instead, we estimated an equivalent effective heat transfer coefficient for the holder contact and applied it as a distributed sink on the modeled radial boundary using an area equivalence argument. This approach is intentionally conservative, in the experiment the holder clamps primarily at the bottom edge rather than uniformly along the sidewall, so distributing the loss over the full radial surface tends to overestimate its influence while still providing a useful bound. We estimate a local contact conductance for holder material thermal conductivity $k_{\mathrm{holder}}$ across an effective conduction length $L_{\mathrm{eff}}$, using $h_{\mathrm{local}} \approx k_{\mathrm{holder}}/L_{\mathrm{eff}}$. Taking a conservative zirconia conductivity of $k_{\mathrm{holder}} \approx$ 3 W m$^{-1}$ K$^{-1}$ \cite{verdi_thermal_2021,mistarihi_fabrication_2015} and a small contact area $A_c \approx 5$ mm$^2$, this yields a $h_{\mathrm{local}}$ on the order of 700 W m$^{-2}$ K$^{-1}$ for the assumed $L_{\mathrm{eff}}$. Converting to an area equivalent coefficient on the sidewall of the disk, $h_{eff} \approx (A_c/A_{\mathrm{side}})h_{\mathrm{local}}$, where $A_{\mathrm{side}}$ is the cylindrical side area of a $\approx 20$ mm diameter, 2 mm thick specimen gives $h_{\mathrm{eff,side}} \approx 28$ W m$^{-2}$ $K^{-1}$. Repeating the fitting procedure at the lowest and highest temperatures in the solid molybdenum dataset, inclusion of this effective holder loss again produced only a negligible change in extracted $k$, increasing it by only $0.36 \: \%$ at 1,836 K and by $0.22 \: \%$ at 2,744 K.

To further test sensitivity to radial conductive loss, we applied an intentionally extreme sink of $h = 500$ W m$^{-2}$ K$^{-1}$ on the radial boundary and refit the data. In that case, the fitted $k$ increased $2\: \%$ at 1,836 K and by $0.96\:  \%$ at 2,744 K; however, the corresponding baseline temperature field becomes inconsistent with the experiment. Specifically, the modeled baseline under predicts the measured peak temperature by $21.3\: \%$ at 1,836 K and $7.0 \: \%$ at 2,744 K, where as the nominal model (with no holder loss) matches the baseline temperature within $\mathrm{\approx 10-50 ~K}$. Taken together, these tests indicate that plausible sample holder conductive losses have a negligible impact on the reported SSTDR measured values of $k$ and that the experimental configuration is effectively insensitive to this loss pathway under the present conditions.

\section{Conclusion}

In this work, we advance steady-state temperature differential radiometry (SSTDR) into a more robust platform for high and ultrahigh temperature measurements. The updated implementation combines lock-in infrared thermography with a spatially localized perturbation heater, optionally dual or single sided baseline heating (dual in this work), and in situ hyperspectral pyrometry to determine true temperature and normal spectral emittance (and corresponding absorptivity) at the laser wavelengths. In parallel, we develop a 2D axisymmetric steady-state heat-transfer model for disk geometries with converged nonlinear radiative (and optional convective) boundary conditions and spatially resolved laser heat-flux inputs, together with an analysis workflow that reconciles the quasi-steady modulated experiment with the steady-state forward model used for fitting.

Using this framework, we determine the solid-state thermal conductivity of high-purity molybdenum from 1,500 –3,000 K (to the onset of melting) with excellent agreement to accepted literature values. Because prior datasets in this regime are often inferred indirectly (most commonly from electrical resistivity via the Wiedemann–Franz law), these measurements provide a high confidence, non-contact benchmark for validating transport models without requiring electrical conductivity relations. More broadly, SSTDR is not restricted to electronically conductive materials, enabling direct $k(T)$ measurements where phonons dominate heat transport and resistivity based inference is inapplicable. Although several experimental techniques exist for high-temperature thermophysical characterization, reliable, uncertainty quantified $k(T)$ data above $\mathrm{\sim}$2000 K remain sparse and method dependent. The present work helps close this thermal property gap by providing a reproducible route to high confidence datasets in a regime where boundary condition sensitivity and radiative losses often dominate error.

A further outcome of the platform is simultaneous acquisition of normal spectral emittance $\varepsilon(\lambda,T)$ over 500 – 1000 nm in both solid and liquid states. While these spectra are used operationally to constrain wavelength-specific absorptivity (and thus absorbed power) at the heater and perturbation wavelengths, they also represent valuable standalone radiative property data in a temperature regime where such measurements are comparatively scarce. Co-measurement of $k(T)$, $T_{\text{true}}$, and phase-dependent $\varepsilon(\lambda,T)$ under the same controlled thermal conditions strengthens the metrological utility of SSTDR and can inform broader radiative property models near melting. As with other multiwavelength pyrometric approaches, however, simultaneous inference of temperature and emittance is not fully unconstrained and is most robust when supplemented by prior knowledge or physically motivated assumptions about the spectral form of emittance (e.g., for pure metals, oxidized surfaces, or ceramics), as widely noted in the literature.

We further perform comprehensive uncertainty bounding studies to quantify measurement fidelity and model robustness. Relative to the initial SSTDR implementation, the present approach reduces uncertainty to $2\sigma$ values of $\mathrm{7.9-11~\%}$ for molybdenum $k$, with uncertainty budget dominated by the measured $\Delta T$ signal and absorbed perturbation power. Sensitivity and bounding studies show that, for the present geometries and operating conditions, the perturbative response is strongly conduction dominated: $\Delta T$ remains highly sensitive to $k$ with weak emittance dependence, and plausible convective and sample holder losses produce negligible changes in fitted $k$. We emphasize that this behavior will not necessarily hold across all SSTDR applications; larger perturbation footprints, thicker specimens, lower thermal conductivity materials, or higher operating temperatures will increase the relative importance of radiation and other boundary condition effects. While this conduction dominant behavior holds for the present geometries and material, the same modeling, sensitivity, and bounding workflow developed here is general and can be applied directly to other SSTDR configurations to quantify when radiation and parasitic losses become important and to maintain traceable uncertainty across materials and temperature regimes.

Looking forward, the combination of localized differential heating, in situ radiometric property assessment, and uncertainty quantified model inversion establishes SSTDR as a robust, traceable framework for high-confidence, direct thermal conductivity measurements in regimes where conventional contact approaches become increasingly uncertain. By eliminating the need for resistivity relations (and, in steady-state form, avoiding reliance on heat capacity inputs), the method provides a versatile path to extend ultrahigh-temperature $k(T)$ benchmarks to ceramics and compositionally complex materials relevant to hypersonics, fusion or fission, and laser-based manufacturing. In parallel, co-measured phase-dependent normal spectral emittance provides radiative property data that can help inform optical thermal modeling near melting and in extreme heat flux environments, where measurement conditions and material behavior permit reliable inference.

\printcredits

\section*{Data availability}

Data will be made available on request.

\section*{Declaration of competing interest}

The authors declare that they have no known competing
financial interests or personal relationships that could have
appeared to influence the work reported in this paper.

\section*{Acknowledgments}

We appreciate support from the Office of Naval Research, Grant No. N00014-26-1-2076 The authors gratefully acknowledge Ralph Felice of FAR Associates\textsuperscript{TM} for insightful technical discussions on the use of spectropyometry.

%\bibliographystyle{cas-model2-names}
\bibliographystyle{model1a-num-names}

\bibliography{cas-refs}


\end{document}